\documentclass[12pt]{article}
\usepackage{amsmath,graphicx}
\usepackage{tabularx}
\usepackage{psfrag}
\usepackage{xcolor}
\usepackage{amsfonts}
\usepackage{amssymb}
\usepackage{amsthm}
\usepackage[ruled,lined,linesnumbered,boxed]{algorithm2e}
\usepackage{algorithmic}
\usepackage{booktabs}
\usepackage{caption}
\usepackage{subcaption}
\usepackage{multirow}
\usepackage{tikz}
\usepackage{colortbl}
\theoremstyle{definition}

\graphicspath{{figures/}}

\begin{document}

\title{Alternative design of DeepPDNet in the context of image restoration}


\author{Mingyuan Jiu \thanks{School of Computer and Artificial Intelligence, Zhengzhou University, Zhengzhou, 450001, China. Email: iemyjiu@zzu.edu.cn} \and
Nelly Pustelnik \thanks{Univ Lyon, Ens de Lyon, Univ Lyon 1, CNRS, Laboratoire de Physique, Lyon, 69342, France and with ISPGroup/ICTEAM, UCLouvain, Belgium. Email: nelly.pustelnik@ens-lyon.fr}
}


\maketitle

\begin{abstract}
This work designs an image restoration deep network relying on unfolded Chambolle-Pock primal-dual iterations. Each layer of our network is built from Chambolle-Pock iterations when specified for minimizing a sum of a $\ell_2$-norm data-term and an analysis sparse prior. The parameters of our network are the step-sizes of the Chambolle-Pock scheme and the linear operator involved in sparsity-based penalization, including implicitly the regularization parameter. A backpropagation procedure is fully described. Preliminary experiments illustrate the good behavior of such a deep primal-dual network in the context of image restoration on BSD68 database. 
\end{abstract}

\section{Introduction}
\label{sec:intro}

Image restoration is a well-studied image processing task where there are still remaining obstacles to be raised, among them,  the design of faster algorithms to accurately restore very large-scale images and the automatic adjustment of hyperparameters.

During the past twenty years, major improvements were made possible in this field with the rise of proximal methods, especially primal-dual proximal methods, allowing to handle with analysis sparse penalization in variational formulations and that drastically improved the quality of the restoration (e.g. total-variation~\cite{RudinOsher92PhysD}, sparse penalization applied on frame coefficients~\cite{Danielyan2012}, non-local TV~\cite{ChierchiaNelly2014tip, li2017:non_local_TV}). However, the question of the hyperparameters selection, which has a major impact on the restoration result, stays a challenging task (see a contrario SURE-based approaches \cite{Stein_C_1981_j-annals-statistics_estimation_mmnd,Deledalle_C_2014_Stein_ugerms}). 

A more recent alternative to nonsmooth optimization relies on supervised neural network learning. The design can be made empirically, with a Plug-and-play (PnP) strategy,  or in an unrolled/unfolded fashion (see~\cite{mccann2017convolutional, Lucas_A_2018_j-ieee-spm, Gilton2019, Ravishankar2020} for review papers). The first class of approaches leads to good performance but suffers from its ``black-box''  lack of interpretation. The second and third ones appear to be more intuitive for experts in the field of image restoration because their architectures rely on the combination of an objective function and an algorithm, and may benefit from the inverse problem literature knowledge. The pioneering work of unrolled algorithm for image analysis is the work by Gregor and LeCun~\cite{Lecun_2010_icml_lfasc}  in a context of sparse coding  relying on forward-backward iterations. A large number of contributions were then related to PnP strategy into ADMM iterations \cite{Venkatakrishnan_S_2013_p-ieee-gcsip_plug_npp,Rond_A_2016_j-vcir_poisson_ipp,Chan_S_2016_j-ieee-tci_Plug-npa,Zhang_K_2017_p-ccvpr_lea_dcn,Gavaskar_R_2019_j-ieee-spl_proof_fpc,Wei_K_2020_p-icml_tuning_pnp, ZhangKaiCVPR2020, NingJSTSP2021} or into primal-dual proximal (PDGH) splitting techniques \cite{Ono_S_2017_j-ieee-spl_primal_pnp,Meinhardt_T_2017_p-ieee-iccv_learning_pou}. Unfolded proximal interior point iterations have been studied in~\cite{Bertocchi_C_2020_j-ip_deep_upp}, and more recently, several unfolded proximal primal-dual iterations have been proposed such as  in \cite{Adler_J_2018_j-ieee-tmi_lea_pdr, JiuPustelnik2020}.  

\noindent \textbf{Context} -- Similarly as in \cite{JiuPustelnik2020}, this work focuses on a restoration problem where 
\begin{equation}
\label{eq:model}
{\textrm z} =  A\overline{\textrm x} + \varepsilon
\end{equation}  involving a linear degradation $ A\in \mathbb{R}^{M\times N}$ and  a Gaussian random degradation $\varepsilon\sim \mathcal{N}(0,\alpha^2 \mathbb{I}_M)$ with a  standard deviation $\alpha$, and where the neural network architecture is built from unrolled iterations of Condat-V\~u iterations \cite{condat_primal-dual_2013, Vu2013} associated to the minimization formulation of this form:
\begin{equation}\label{eq:general_prob}
\widehat{\textrm x}_\lambda\in \underset{\textrm x \in \mathbb{R}^N}{\mathrm{Argmin}} \frac{1}{2}\Vert  A \textrm x - \textrm z \Vert_2^2 +  \lambda g( D\textrm x),
\end{equation}
where $D\in \mathbb{R}^{P\times N}$ denotes the analysis sparsifying transform  and $g\colon \mathbb{R}^P \to ]-\infty,+\infty]$ is typically a proper convex lower-semi continuous function, which models a sparse penalization~\cite{Pustelnik_N_20016_j-w-enc-eee_wav_bid,Bach2012} (i.e. a $\ell_1$-norm or a $\ell_{1,2}$-norm that favors coupling between coefficients), and $\lambda>0$ stands for the regularization parameter acting as a trade-off between the data-fidelity term and the penalization.

\noindent \textbf{Contributions and outline} -- Considering Condat-V\~u iterations, the data-fidelity term can be either activated through a gradient step or through a proximal step leading to two different networks. In \cite{JiuPustelnik2020}, the activation as a gradient step has been explored while the contribution of this work focuses on its proximal activation, related to Chambolle-Pock iterations. Our contribution aims first to provide the associated neural network architecture (cf. Section~\ref{sec:cp}),  to derive a backpropagation procedure in order to learn the algorithmic parameter step-sizes and the linear operator $D$ (and implicitly the regularization parameter $\lambda$) as described in Section~\ref{sec:cpdeeppdnet}, and finally to illustrate the good behaviour of the proposed Proximal activation (PA) DeepPDNet in the context of image restoration on BSD68 database (cf. Section~\ref{sec:num}).

\section{Proximal activation of DeepPDNet}
\label{sec:cp}
The design of our neural network  relies on a criterion based on a reformulation of \eqref{eq:general_prob} in order to facilitate the joint learning of $\lambda$ and $D$, which writes
\begin{equation}\label{eq:basic_prob}
\widehat{\mathrm x}\in \underset{\mathrm x \in \mathbb{R}^N}{\mathrm{Argmin}} \;\frac{1}{2}\Vert  A \mathrm x - \mathrm z \Vert_{2}^2+ h(L\mathrm x).
\end{equation}
where $L\in \mathbb{R}^{P\times N}$ implicitly combines the information of $\lambda$ and $D$ and where $h$ is a convex, lower semi-continuous, and proper function from $\mathbb{R}^P$ to $]-\infty,+\infty]$. 

\subsection{Chambolle-Pock iterations}
The Chambolle-Pock iterations~\cite{ChambollePock2011},  in the specific context of \eqref{eq:basic_prob},  reads, for every $k>0$,
\begin{equation}
\label{eq:FPPD}
\begin{cases}
\textrm{y}^{[k+1]} &= \textrm{prox}_{\sigma h^{*}} \big(\textrm{y}^{[k]} + \sigma  L  {\overline{\textrm{x}}}^{[k]} \big)\\
\textrm{x}^{[k+1]} &= (\tau A^*A + \textrm{I})^{-1}(\tau A^*\textrm{z} + \textrm{x}^{[k]} - \tau L^* \textrm{y}^{[k+1]} ) \\
\overline{\textrm{x}}^{[k+1]} &= \textrm{x}^{[k+1]} +\theta(\textrm{x}^{[k+1]}  - \textrm{x}^{[k]} )
\end{cases}
\end{equation}
where $\theta$, $\tau$ and $\sigma$ are algorithmic  parameters and  where $\textrm{prox}$ denotes the proximity operator \cite{bauschke_convex_2017} which is defined for a proper convex lower semi-continuous function $f\colon \mathcal{H}\to ]-\infty,+\infty]$, when $ \mathcal{H}$ models a real Hibert space, as for every $\textrm{x} \in \mathcal{H}$, $\text{prox}_f(\textrm{x})=\underset{\textrm{y}\in \mathcal{H}}{\text{argmin}}\frac{1}{2}\Vert \textrm{y}-\textrm{x}\Vert_2^2+f(\textrm{y})$.
$h^{*}$ is the Fenchel-Rockafellar conjugate function of $h$  and we recall that $\text{prox}_{h^*}$ can be easily computed from $\text{prox}_{h}$  using Moreau identity $\text{prox}_{\sigma h^*}(y)=y - \sigma \text{prox}_{h/\sigma}(y/\sigma).$ Under technical assumptions, especially involving the choice of the step-size $\tau$ and $\sigma$,  the relaxation parameter $\theta$, and the norm of $L$, the sequence $(\textrm{x}^{[k]})_{k\in\mathbb{N}}$ is insured to converge to $\widehat{\textrm{x}}$. 

\subsection{Reformulation of Chambolle-Pock iterations}
As a preliminary step to understand our neural network architecture, we propose to rewrite Chambolle-Pock iterations~\eqref{eq:FPPD} when $\theta=0$ as it follows:
\begin{equation}
\label{eq:FPPD2}
\begin{cases}
\textrm{y}^{[k+1]} &= \textrm{prox}_{\sigma h^{*}} \big(\textrm{y}^{[k]} + \sigma  L  \textrm{x}^{[k]} \big)\\
\textrm{x}^{[k+1]} &= (\tau A^*A + \textrm{I})^{-1}(\tau A^*\textrm{z} + \textrm{x}^{[k]} - \tau L^*(\textrm{y}^{[k]} + \sigma  L  \textrm{x}^{[k]})\\
& +  \sigma \tau L^* \textrm{prox}_{\sigma^{-1} h} \big(\sigma^{-1}\textrm{y}^{[k]} + L  \textrm{x}^{[k]}) \big) 
\end{cases}
\end{equation}
or equivalently,
\begin{equation}
\begin{cases}
\textrm{h}_1^{[k+1]} = \tau A^* \textrm{z} + \textrm{x}^{[k]} - \tau L^* (\textrm{y}^{[k]}+ \sigma L \textrm{x}^{[k]}) \\
\textrm{h}_2^{[k+1]} = \textrm{prox}_{\sigma^{-1}h} \big(\sigma^{-1} \textrm{y}^{[k]}+L\textrm{x}^{[k]} \big) \\
\textrm{h}_3^{[k+1]} = \textrm{prox}_{\sigma\lambda h^{*}} \big(\textrm{y}^{[k]} + \sigma  L  \textrm{x}^{[k]}) \big)\\
\textrm{x}^{[k+1]} = (\tau A^*A + \textrm{Id})^{-1} (\textrm{h}_1^{[k+1]} + \sigma \tau L^{*} \textrm{h}_2^{[k+1]}) \\
\textrm{y}^{[k+1]} = \textrm{h}_3^{[k+1]}
\end{cases} \label{eq:newFPPD}
\end{equation}
providing a link between Chambolle-Pock iterations and the following feed-forward network architecture:
\begin{equation}
\textrm{u}^{[K]} = H^{[k]} \eta^{[K]} \big( G^{[K]}\ldots G^{[2]}H^{[1]} \eta^{[1]} (G^{[1]}\textrm{u}^{[1]} + b^{[1]}) \ldots + b^{[K]} \big). 
\end{equation}
where  $\textrm{u}^{[k]}=\big( (\textrm{x}^{[k]})^{\top}, (\textrm{y}^{[k]})^{\top} \big)^{\top}$ and having a hidden layer with three nodes  denoted $\textrm{h}^{[k]}_1, \textrm{h}^{[k]}_2, \textrm{h}^{[k]}_3$, and

{\footnotesize{\begin{equation}\begin{cases}
		G^{[k]} =  
		\begin{pmatrix}
		\mathrm{Id} - \tau \sigma L^* L &  -\tau L^*\\
		L & \sigma^{-1} \\
		\sigma L & \mathrm{Id} \\
		\end{pmatrix}\\
		b^{[k]} = 
		\begin{pmatrix}
		\tau A^*\mathrm{z}\\
		0 \\
		0\\
		\end{pmatrix}\\
		\eta^{[k]} =   \begin{pmatrix}
		\textrm{Id} \\
		\textrm{prox}_{\sigma^{-1}h} \\
		\textrm{prox}_{\sigma h^*} \\
		\end{pmatrix} \\
		H^{[k]} =   \begin{pmatrix}
		(\tau A^*A + \textrm{Id})^{-1} & (\tau A^*A + \textrm{Id})^{-1} \sigma \tau L^* & 0 \\
		0 & 0 & \mathrm{Id}\\
		\end{pmatrix}.
		\end{cases} \label{equa:paramdef}\end{equation}}}

\vspace{-0.9cm}
\subsection{Chambolle-Pock DeepPDNet} 
Given the training set $\mathcal{S} = \{ (\overline{\mathrm{x}}_s, \mathrm{z}_s) | s=1, \ldots, I\}$ where $\overline{\mathrm{x}}_s$ is the undegraded image and $\mathrm{z}_s$ is its degraded counterpart following degradation model~\eqref{eq:model}. We build an inverse problem solver $f_{\widehat{\Theta}}$ relying on a neural network architecture involving the parameters $\widehat{\Theta}$. 
The estimation of these parameters relies on the following standard empirical loss:
\begin{equation} \label{eq:newlossk}
\widehat{\Theta} \in \underset{\Theta}{\textrm{Argmin}}\;E(\Theta):= \frac{1}{I}\sum_{s=1}^I \Vert \overline{\mathrm{x}}_s - f_\Theta(A^*\mathrm{z}_s) \Vert_2^2
\end{equation}
where the proposed network writes, for every $\textrm{u}\in \mathbb{R}^N$,
\begin{align}
f_\Theta(\textrm{u}) = H^{[K]} \eta^{[K]} \left( G^{[K]}\ldots H^{[1]} \eta^{[1]} (G^{[1]} \textrm{u} + b^{[1]}) \ldots + b^{[K]} \right) \nonumber
\end{align}
with $\widehat{\Theta} = \{\widehat{\sigma}^{[k]}, \widehat{\tau}^{[k]}, \widehat{L}^{[k]}\}_{1\leq k \leq K}$ and 
{\footnotesize \begin{equation} 
	\begin{cases}
	G^{[k]} =  
	\begin{pmatrix}
	\mathrm{Id} - \tau^{[k]} \sigma^{[k]} L^{[k]*} L^{[k]} &  -\tau^{[k]} \big(L^{[k]}\big)^*\\
	L^{[k]} & (\sigma^{[k]})^{-1} \\
	\sigma^{[k]} L^{[k]} & \mathrm{Id} \\
	\end{pmatrix}\\
	b^{[k]} = 
	\begin{pmatrix}
	\tau^{[k]} A^*\mathrm{z}_s\\
	0 \\
	0\\
	\end{pmatrix}\\
	\eta^{[k]} =   \begin{pmatrix}
	\textrm{Id} \\
	\textrm{prox}_{h/\sigma^{[k]}} \\
	\textrm{prox}_{\sigma^{[k]} h^*} \\
	\end{pmatrix} \\
	H^{[k]} =   \begin{pmatrix}
	(\tau^{[k]} A^*A  + \textrm{Id})^{-1} &  \sigma^{[k]} \tau^{[k]}(\tau^{[k]} A^*A  + \textrm{Id})^{-1} \big(L^{[k]}\big)^* & 0 \\
	0 & 0 & \mathrm{Id}\\
	\end{pmatrix} 
	\end{cases} \label{eq:newstructure}
	\end{equation}}
and the first and last layers are:
{\footnotesize \begin{equation} 
	\begin{cases}
	G^{[1]} =  
	\begin{pmatrix}
	\mathrm{Id} - \tau^{[1]} \sigma^{[1]} \big(L^{[1]}\big)^* L^{[1]} \\
	L^{[1]} \\
	\sigma^{[1]} L^{[1]} \\
	\end{pmatrix}\\
	G^{[K]} =  
	\begin{pmatrix}
	\mathrm{Id} - \tau^{[K]} \sigma^{[K]} L^{[K]*} L^{[K]} &  -\tau^{[K]} L^{[K]*}\\
	L^{[K]} & (\sigma^{[K]})^{-1} \\
	\end{pmatrix}\\
	b^{[K]} = 
	\begin{pmatrix}
	\tau^{[K]} A^*\mathrm{z}_s\\
	0 \\
	\end{pmatrix}\\
	\eta^{[K]} =   \begin{pmatrix}
	\textrm{Id} \\
	\textrm{prox}_{h/\sigma^{[K]}} \\
	\end{pmatrix} \\
	H^{[K]} =   \begin{pmatrix}
	(\tau^{[K]} A^*A+ \textrm{Id})^{-1} & \sigma^{[K]} \tau^{[K]} (\tau^{[K]} A^*A + \textrm{Id})^{-1}  L^{[K]*} \\
	\end{pmatrix},	\end{cases} \label{eq:newstructure1}
	\end{equation}}
\noindent The dual variable in the first layer is set to $\textrm{y}^{[1]}=0$, and the last layer is also modified to only output the primal variable, since the ground-truth of dual variable is not known.

\section{Learning procedure}
\label{sec:cpdeeppdnet}
The estimation of  $\widehat{\Theta} = \{\widehat{\sigma}^{[k]}, \widehat{\tau}^{[k]}, \widehat{L}^{[k]}\}_{1\leq k \leq K}$ relies on a gradient based strategy for each parameter and whose iterations are, for every $\ell=0,1,\ldots$, and every layer $k$,
\begin{equation}
\begin{cases}
\tau^{[k]}_{\ell+1} &= \tau^{[k]}_{\ell} - \gamma_\tau \frac{\partial E}{\partial \tau^{[k]}}  \\
\sigma^{[k]}_{\ell+1} &= \sigma^{[k]}_{\ell} - \gamma_\sigma \frac{\partial E}{\partial \sigma^{[k]}}\\
L^{[k]}_{\ell+1} &= L^{[k]}_{\ell} - \gamma_L \frac{\partial E}{\partial L^{[k]}}
\end{cases}  \label{eq:update}
\end{equation}
for some learning rate $\gamma>0$. The computation of $\frac{\partial E}{\partial \theta^{[k]}}$ where $\theta$ models either $\tau$, $\sigma$, or $L$ relies on a backpropagation procedure such as:
\begin{equation}
\label{eq:bp}
\frac{\partial E}{\partial \theta^{[k]}} = \frac{\partial E}{\partial \textrm{u}^{[K]}} \frac{\partial \textrm{u}^{[K]}}{\partial \textrm{u}^{[K-1]}} \ldots \frac{\partial \textrm{u}^{[k+1]}}{\partial \textrm{u}^{[k]}} \frac{\partial \textrm{u}^{[k]}}{\partial \theta^{[k]}}
\end{equation}
We set
\vspace{-0.7cm}

$$
\begin{cases}
\textrm{v}^{[k]} = G^{[k]}\textrm{u}^{[k-1]} + b^{[k]}\\
\textrm{w}^{[k]} = \eta^{[k]} (\textrm{v}^{[k]})\\
\textrm{u}^{[k]} =   H^{[k]} \textrm{w}^{[k]}
\end{cases}
$$ 
leading to
\begin{equation}
\label{eq:eq10}
\frac{\partial \textrm{u}^{[k]}}{\partial \textrm{u}^{[k-1]}}  = H^{[k]}\frac{d\eta^{[k]}(\textrm{v}^{[k]})}{d\textrm{v}^{[k]} }G^{[k]}
\end{equation}
and
{\footnotesize{\begin{equation}
		\label{eq:eq11}
		\hspace{-0.3cm}\frac{\partial \textrm{u}^{[k]}}{\partial \theta^{[k]}} =
		H^{[k]} \underbrace{\Bigg( \frac{\partial \eta^{[k]}(\mathrm{v}^{[k]})}{\partial \mathrm{v}^{[k]}} \left( \frac{\partial G^{[k]}}{\partial \theta^{[k]}} \textrm{u}^{[k-1]} + \frac{\partial b^{[k]}}{\partial \theta^{[k]}} \right)+ \frac{\partial \eta^{[k]}( \mathrm{v}^{[k]})}{\partial \theta^{[k]}}\Bigg)}_{ \frac{\partial \textrm{w}^{[k]} }{\partial \theta^{[k]}}}
		+ \frac{\partial H^{[k]}}{\partial \theta^{[k]}}\textrm{w}^{[k]} 
		\end{equation}}}
The learning procedure is summarized in Algorithm~\ref{algo:backwardk}.

\begin{algorithm}
		
		\small
		\caption{Learning algorithm for PA-DeepPDNet} \label{algo:backwardk}
		\BlankLine
		\KwIn{Set $\widehat{\Theta}_0 = \{\widehat{\sigma}_0^{[k]}, \widehat{\tau}_0^{[k]}, \widehat{L}_0^{[k]}\}_{1\leq k \leq K}$.\\
			$\qquad$ Set $G_0^{[k]}$, $b_0^{[k]}$, $\eta_0^{[k]}$, $H_0^{[k]}$ according to \eqref{eq:newstructure} and \eqref{eq:newstructure1}.\\
			$\qquad$  Set $\gamma_\theta>0$, where $\theta$ either denotes $\sigma$, $\tau$, or $L$.}
		\KwData{Set $\textrm{u}^{[1]}_s = A^*\textrm{z}_s, s=\{1, \ldots, I\}$}
		\For{$\ell = 0, \ldots, \mathrm{itermax}$}{
			Select one (or several) training sample $\textrm{u}^{[1]}_s = A^*\textrm{z}_s$.\\
			Compute $\textrm{u}^{[K]}_s = f_{\widehat{\Theta}_{\ell}}(\textrm{u}^{[1]}_s)$.\\
			$t \leftarrow \frac{dE}{d \textrm{u}^{[K]}_s}$ according to Eq.~\eqref{eq:gradient1}.\\
			\For{$k =K, \ldots, 1$}{
				Compute $\frac{d \textrm{u}^{[k]}_s}{d \theta_\ell^{[k]}}$ according to \eqref{eq:eq11}.\\
				Compute $\frac{\partial E}{\partial \theta_\ell^{[k]}}$ according to \eqref{eq:bp}:
				$ \frac{d E}{d \theta_\ell^{[k]}} \leftarrow t \times \frac{d \textrm{u}^{[k]}_s}{d \theta_\ell^{[k]}}$.\\		
				Backpropagate considering  \eqref{eq:eq10}:
				$t \leftarrow t\times \frac{d \textrm{u}^{[k]}_s}{d \textrm{u}^{[k-1]}_s}$.\\
			}
			For every $k$, update the parameter $\theta^{[k]}$:\\$\theta^{[k]}_{\ell+1} \leftarrow  \theta^{[k]}_{\ell} - \gamma_\theta \frac{\partial E}{\partial \theta_\ell^{[k]}}$. \\
		}
\end{algorithm} 

We give a closed form for each involved derivation in Algorithm~\ref{algo:backwardk}:\\
\noindent $\bullet$ The error of loss $E$ w.r.t. $\textrm{u}^{[K]}_s$ is
\begin{equation} \label{eq:gradient1}
\frac{\partial E  }{\partial \textrm{u}^{[K]}} =   \frac{2}{I}(\textrm{u}^{[K]} - \overline{\textrm{x}}).
\end{equation}

\noindent $\bullet$ In the specific case where $h= \Vert \cdot \Vert_1$, the error of the hidden variable $\textrm{w}^{[k]}$ w.r.t. $\textrm{v}^{[k]} = (\textrm{v}^{[k]}_{1}, \textrm{v}^{[k]}_{2}, \textrm{v}^{[k]}_{3})$ is defined as:
\begin{equation}
\frac{\partial \eta^{[k]}(\mathrm{v}^{[k]})}{\partial \mathrm{v}^{[k]}}  = (\textrm{r}_1^\top,\textrm{r}_2^\top, \textrm{r}_3^\top)^\top {\in \mathbb{R}^{N+2P}}
\end{equation}
\noindent where $\textrm{r}_1 =(1,1,\ldots,1)\in \mathbb{R}^N$, and for every $p\in\{1,\ldots,P\}$,
\begin{equation}
\footnotesize
\textrm{r}_{2,p} = \begin{cases}
1 &  \mbox{if}\;\vert \textrm{v}^{[k]}_{2,p}\vert > \frac{1}{\sigma^{[k]}} \\
0 &  \mbox{if}\;\vert \textrm{v}^{[k]}_{2,p}\vert < \frac{1}{\sigma^{[k]}} \\
[0,1] &\mbox{if}\; \textrm{v}^{[k]}_{2,p} =\pm\frac{1}{\sigma^{[k]}},
\end{cases}
\ \ \ 
\textrm{r}_{3,p} = \begin{cases}
0 &  \mbox{if}\;\vert \textrm{v}^{[k]}_{3,p}\vert >1 \\
1 &  \mbox{if}\;\vert \textrm{v}^{[k]}_{3,p}\vert <1 \\
[0,1] &\mbox{if}\; \textrm{v}^{[k]}_{3,p} =\pm1.
\end{cases} \label{eq:subgradproximall1}
\end{equation}

\noindent $\bullet$ Since $\textrm{w}^{[k]}=(\textrm{w}^{[k]}_{1}, \textrm{w}^{[k]}_{2}, \textrm{w}^{[k]}_{3})$,  are respectively the identity, the proximity operator of $\ell_1$-norm and the proximity operator of the conjugate of the $\ell_1$-norm (corresponding to $\textrm{prox}_{h/\sigma^{[k]}}$ and $\textrm{prox}_{\sigma^{[k]} h^*}$), so their sub-differential w.r.t. $\sigma_\ell^{[k]}$ are:
\begin{equation}
\frac{\partial \textrm w^{[k]}_{1}}{\partial \sigma^{[k]}_{\ell}}=0, \ \ \ 	\frac{\partial \textrm w^{[k]}_{3}}{\partial \sigma^{[k]}_{\ell}}= 0,	
\end{equation}
\begin{equation}
\frac{\partial \textrm w^{[k]}_{2,{p}}}{\partial \sigma^{[k]}_{\ell}}  = \begin{cases} 
0 & \vert \textrm{v}^{[k]}_{2,p}\vert < \frac{1}{\sigma^{[k]}}  \\
\frac{1}{\sigma^{[k]2}} & \textrm{v}^{[k]}_{2,p} > \frac{1}{\sigma^{[k]}} \\
-\frac{1}{\sigma^{[k]2}} & \textrm{v}^{[k]}_{2,p} < -\frac{1}{\sigma^{[k]}} \\
[0, \frac{1}{\sigma^{[k]2}}] & \textrm{v}^{[k]}_{2,p} = \frac{1}{\sigma^{[k]}} \\
[-\frac{1}{\sigma^{[k]2}}, 0] & \textrm{v}^{[k]}_{2,p} = -\frac{1}{\sigma^{[k]}} \\	 
\end{cases} 
\end{equation}

\noindent $\bullet$ The remaining gradient involved in Eq.~\eqref{eq:eq11} are:
{\footnotesize \begin{align} \label{eq:derivativelist}
	\frac{\partial b^{[k]}_{\ell}}{\partial \tau^{[k]}_{\ell}} & = \begin{pmatrix}
	A^*\textrm{z}_s\\
	0 \\
	0 
	\end{pmatrix}  \qquad
	\frac{\partial b^{[k]}_{\ell}}{\partial \sigma^{[k]}_{\ell}}  = 0 \qquad 
	\frac{\partial b^{[k]}_{\ell}}{\partial L^{[k]}_{\ell}}  = 0 \\
	\frac{\partial G^{[k]}_{\ell}}{\partial \tau^{[k]}_{\ell}} & = \begin{pmatrix}
	- \sigma^{[k]}_{\ell} L^{[k]*}_{\ell} L^{[k]}_{\ell}& - L^{[k]*}_{\ell} \\
	0 & 0 \\
	0 & 0 
	\end{pmatrix} \\
	\frac{\partial G^{[k]}_{\ell}}{\partial \sigma^{[k]}_{\ell}} & = \begin{pmatrix}
	-\tau^{[k]}_{\ell}L^{[k]*}_{\ell}L^{[k]}_{\ell} & 0 \\
	0 & - (\sigma^{[k]}_{\ell})^{-2} \\
	L^{[k]}_{\ell} & 0
	\end{pmatrix} \\
	\frac{\partial G^{[k]}_{\ell}}{\partial L^{[k]}_{\ell}} & = \begin{pmatrix}
	-2\tau^{[k]}_{\ell}\sigma^{[k]}_{\ell}L^{[k]*}_{\ell} & - \tau^{[k]}_{\ell} \\
	1 & 0 \\
	\sigma^{[k]}_{\ell} & 0
	\end{pmatrix} \\	
	\frac{\partial H^{[k]}_{\ell}}{\partial \sigma^{[k]}_{\ell}} & = \begin{pmatrix}
	0 &  \tau^{[k]}_{\ell} F^{-1}(\tau^{[k]}_{\ell} \Lambda^2  + \textrm{Id})^{-1}F L^{[k]*}_{\ell} & 0 \\
	0 & 0 & 0\\
	\end{pmatrix} \\
	\frac{\partial H^{[k]}_{\ell}}{\partial L^{[k]}_{\ell}} & = \begin{pmatrix}
	0 & \sigma^{[k]}_{\ell} \tau^{[k]}_{\ell} F^{-1}(\tau^{[k]}_{\ell} \Lambda^2  + \textrm{Id})^{-1}F& 0 \\
	0 & 0 & 0 \\
	\end{pmatrix} \\
	\frac{\partial H^{[k]}_{\ell}}{\partial \tau^{[k]}_{\ell}} & = \begin{pmatrix}
	F^{-1} B F & F^{-1} C F\sigma^{[k]} L^{[k]*} & 0 \\
	0 & 0 & 0
	\end{pmatrix} 	
	\end{align}}
\noindent where the last expression is obtained using the specific property of circulant matrices $A = F^*\Lambda F$, leading to
$(\tau A^*A + \textrm{I})^{-1}= F^{-1}(\tau \Lambda^2  + \textrm{I})^{-1}F$. Consequently, $B$ and $C$ may be defined as diagonal matrices where  the diagonal elements are $B_{ii}=\frac{-\Lambda_{ii}^2}{(\tau^{[k]}_{[l]}\Lambda_{ii}^2+1)^2}$ and $C_{ii}=\frac{1}{(\tau^{[k]}_{\ell}\Lambda_{ii}^2+1)^2}$.

\begin{figure*}[t]
	\centering
	\begin{tabular}{cccc}		
		\includegraphics[height=0.13\linewidth]{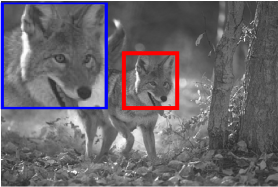} &
		\includegraphics[height=0.13\linewidth]{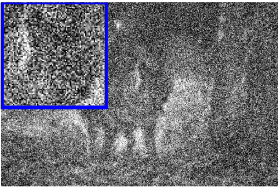} &
		\includegraphics[height=0.13\linewidth]{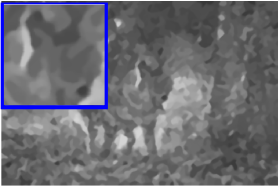} & 
		\includegraphics[height=0.13\linewidth]{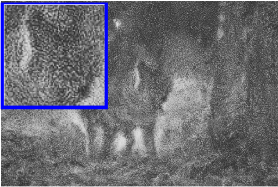} \\
		& {$z$: 11.58 dB} & {NLTV \cite{ChierchiaNelly2014tip}: 23.76 dB} & {MWCNN \cite{LieZhang-wmcnn_cvpr18}: 17.75 dB} \\	
		&
		\includegraphics[height=0.13\linewidth]{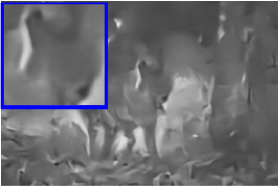} &
		\includegraphics[height=0.13\linewidth]{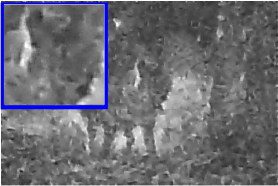} &
		\includegraphics[height=0.13\linewidth]{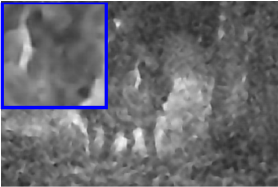}	\\
		& {IRCNN \cite{Zhang_K_2017_p-ccvpr_lea_dcn}: 23.80 dB} & {DeepPDNet \cite{JiuPustelnik2020}: 22.52 dB} & {PA-DeepPDNet: 23.96 dB} \\											
	\end{tabular}
	\caption{\small Visual comparisons on BSD68 dataset for different methods with a uniform $5\times5$ blur and a Gaussian noise with $\alpha=75$. The images respectively correspond to the clean image, the degraded one $\textrm{z}$, the restored ones by NLTV, MWCNN, IRCNN and the DeepPDNet and the proposed PA-DeepPDNet ($K=10$), as well as the PSNR below the image. The region in the blue box are the zoomed region in the red box. } \label{fig:bsdinstance}
	\vspace{-0.5cm}
\end{figure*}

\section{Numerical experiments} \label{sec:num}

\noindent\textbf{Database} -- In this section, we evaluate the performance of the proposed network to image restoration task on the well-known gray version of BSD68 database~\cite{Roth2009Fields}, which contains 68 natural images of size $321\times481$ extracted from Berkeley dataset~\cite{Roth2009Fields}. For the training, we  follow~\cite{Chen2017Trainable}, and use 400 images of size $180\times180$ from the Berkeley dataset which does not contain the 68 used for testing.

A patch-based strategy is adopted for the training procedure. We randomly collect a set of 260000 patches of size $10\times10$ from the training dataset described previously. ADAM strategy is used for the learning \cite{Kingma_D_2015_adam_iclr}. This learned local network is then slid on the test images to obtain the restored image and evaluate the performance. 

\noindent\textbf{Performance Assessment} --The performance are evaluated in terms of PSNR (i.e.~Peak Signal-to-Noise Ratio).
Four different degradation scenarios are considered: $3\times3$ and $5\times5$ uniform blur and additional noises with standard deviation $\alpha=25$, 50, and 75.

We compare the proposed PA-DeepPDNet (for Proximal Activation DeepPDNet) with the standard TV~\cite{RudinOsher92PhysD}, NLTV~\cite{ChierchiaNelly2014tip}, EPLL~\cite{Zoran_D_2017_p-iccv_fro_lmn} restoration procedures, with deep learning procedures MWCNN~\cite{LieZhang-wmcnn_cvpr18}, IRCNN~\cite{Zhang_K_2017_p-ccvpr_lea_dcn}, and our previous DeepPDNet~\cite{JiuPustelnik2020} built from the gradient activation of the data-fidelity term in the Condat-V\~u iterations. For the standard approaches the regularization parameter is set by cross-validation on set12 dataset \cite{Zhang2017}.

\noindent\textbf{Architecture specificities} -- Inspired from the feature design in~\cite{JiuPustelnik2020}, we choose a mixture of global and local sparse features to construct the $L^{[k]}$, where each row models either a global (dense) or local pattern (convolutional). We consider the design named \textit{f5s2n30 + f7s3n30 +  f10s10n30}  leading to $L^{[k]}\in \mathbb{R}^{420\times100}$. The $L^{[k]}$ is randomly initialized by a normal distribution with standard deviation of $10^{-2}$. 

We build a network with $K=10$ layers and each layer is initialized with the same parameters $G_0^{[k]}$, $\eta_0^{[k]}$, $b_0^{[k]}$ and $H_0^{[k]}$ according to Eq.~\eqref{eq:newstructure} and \eqref{eq:newstructure1}, where $\tau_{[0]}^{[k]}$ and $\sigma_{[0]}^{[k]}$ and $L_{[0]}^{[k]}$ are set to satisfy the constraint: $\tau_{[0]}^{[k]}\sigma_{[0]}^{[k]}\Vert L_{[0]}^{[k]}\Vert^2<1$. Once the network is initialized, the parameters are updated by the proposed learning algorithm described in Sec.~\ref{sec:cpdeeppdnet}. In the learning procedure, we adopt a mini-batch strategy with 200 samples for each batch and $8\times10^5$ maximal iterations.

\noindent \noindent\textbf{Complexity analysis and comparison to DeepPDNet} -- When $P \gg N$ (as considered in this experimental section), the complexity cost for one forward layer operation can be approximated as $P^2$ for DeepPDNet and $2P^2$ for PA-DeepPDNet. Such complexity is confirmed by the experiment when the network forward procedures of the learned models are evaluated per patch and lead to an average running time of $0.0022$ (sec.) with DeepPDNet (20 layers) and $0.0023$ (sec.) with PA-DeepPDNet (10 layers)\footnote{Matlab on a machine with Intel(R) i7-8550U CPU}.

\noindent\textbf{Results} -- The learned local network is slid on the image to obtain the restored images by two fashions: i) the neighboring patches are restored independently (cf. PA-DeepPDNet-Independent); ii) the neighboring patches have overlaps and a final average result is computed for each pixel (cf. PA-DeepPDNet-Averaged), always leading to better performance. The comparison results on the test set are shown in Tab.~\ref{tab:bsdfinalcomp}. It can be seen that: i) the proposed PA-DeepPDNet outperforms the learned DeepPDNet; ii) when the noise standard deviation becomes larger, the PA-DeepPDNet is better than other methods, except when the noise level is $\alpha=25$, where the proposed PA-DeepPDNet is 0.3dB lower than IRCNN. A reasonable explanation comes from three possible reasons: i) the choice $\theta=0$ in Eq.~\eqref{eq:FPPD}, which has been made in order to facilitate the learning but maybe at the price of a lack of efficiency; ii) the receptive field of the current local features, which is relatively small and limited by the patch size ($10\times10$); a backbone of off-shelf CNN module can be further integrated into the framework to improve the performance, especially when $\alpha=25$; iii) when the noise level overwhelms  the blur (especially when $\alpha=75$), the prior knowledge about the blur ($A$ in the Eq.~(4) in the revision) takes more important role on the restoration and it finally guides a better solution in the learning iterations. A deeper analysis will be done in future work. Fig.~\ref{fig:bsdinstance} displays examples of original images, degraded images, and restored ones by the different methods.

{{\begin{table}[h]
			\centering
			\resizebox{\linewidth}{!}{
				\begin{tabular}{|c|c|c|c|c|c|c|c|c|}
					\hline
					\multirow{2}{*}{Method} & \multicolumn{3}{c|}{Blur filter $3\times3$} & \multicolumn{3}{c|}{Blur filter $5\times5$}  \\
					\cline{2-7}
					& $\alpha=25$ & $\alpha=50$ & $\alpha=75$ & $\alpha=25$ & $\alpha=50$ & $\alpha=75$ \\     
					\hline
					TV~\cite{RudinOsher92PhysD} & 25.31 & 23.30 & 21.81 & 24.18 & 23.02 & 21.26 \\
					NLTV~\cite{ChierchiaNelly2014tip} & 25.69 & 23.58 & 21.82 & 24.43 & 23.28 & 21.65 \\
					EPLL~\cite{Zoran_D_2017_p-iccv_fro_lmn} & 25.59 & 23.73 & 20.75 & 24.42 & 23.02 & 20.72 \\
					\hline
					MWCNN~\cite{LieZhang-wmcnn_cvpr18} & 25.94 & 24.00 & 17.87 & 24.29 & 23.05 & 17.49 \\
					IRCNN~\cite{Zhang_K_2017_p-ccvpr_lea_dcn} & \textbf{26.36} & 23.63 & 21.92 & \textbf{25.01} & 22.99 & 21.44 \\		
					Learned DeepPDNet~\cite{JiuPustelnik2020} & 25.75 & 23.56 & 21.06 & 23.55 & 22.60 & 20.77 \\
					\hline		
					PA-DeepPDNet (Independent) & 25.76 & 23.84 & 22.63 & 24.57 & 23.09 & 22.26 \\				
					PA-DeepPDNet (Averaged) & 26.02 & \textbf{24.09} & \textbf{22.87} & 24.69 & \textbf{23.29} & \textbf{22.36} \\							
					\hline
				\end{tabular}
			}
			\vspace{-0.2cm}
			\caption{\small Comparison PSNR results of different methods on the BSD68 dataset from different degradation configurations.} \label{tab:bsdfinalcomp}
\end{table}}}

\vspace{-0.5cm}

\section{Conclusion}
In this work, we propose Proximal alternative to our DeepPDNet. The backpropagation procedure is fully detailed allowing to reproduce easily this learning-based restoration strategy. We experimented with the proposed approach on BSD68 dataset, and obtain competitive results that are encouraging as being comparable to state-of-the-art results. However, in future work, on one hand, an end-to-end the CNN network can be combined into framework to further improve the performance; on the other hand, deeper analysis on complete reformulation~\eqref{eq:FPPD}, including the learning of $\theta$, will certainly help improve the restoration performances. Additionally, the conclusion between PA-DeepPDNet and the learned DeepPDNet requires a deeper study  as the boundary effects are not dealt similarly.


\end{document}